\documentclass[aps,showpacs,preprintnumbers,amsmath,amssymb,superscriptaddress]{revtex4}

\usepackage{graphicx}
\usepackage{epstopdf}
\usepackage{dcolumn}
\usepackage{amsmath, mathrsfs}
\usepackage{amsmath}
\usepackage{amssymb}
\usepackage{bm}
\usepackage{color}
\usepackage{multirow}
\usepackage{float}
\usepackage[dvipsnames]{xcolor}
\usepackage{hyperref}
\hypersetup{
	colorlinks=true, 
	linkcolor=blue,  
	citecolor=cyan,  
}

\newcommand{\beqn}{\begin{eqnarray}}
\newcommand{\eeqn}{\end{eqnarray}}
\RequirePackage[normalem]{ulem}

\newcommand{\bqn}{\begin{eqnarray}}
\newcommand{\eqn}{\end{eqnarray}}

\newcommand{\lb}{\label}

\begin{document}
\baselineskip=0.8 cm	
\title{Signatures from the observed jet power and the radiative efficiency for rotating black holes in loop quantum gravity
 }
\author{Zhengwei Cheng$^{1}$, Songbai Chen$^{1,2}$\footnote{csb3752@hunnu.edu.cn}, Jiliang Jing$^{1,2}$
\footnote{jljing@hunnu.edu.cn}}
\affiliation{ $^1$Department of Physics, Institute of Interdisciplinary Studies, Key Laboratory of Low Dimensional Quantum Structures
    and Quantum Control of Ministry of Education, Synergetic Innovation Center for Quantum Effects and Applications, Hunan
    Normal University,  Changsha, Hunan 410081, People's Republic of China
    \\
    $ ^2$Center for Gravitation and Cosmology, College of Physical Science and Technology, Yangzhou University, Yangzhou 225009, People's Republic of China}

\date{\today}

\begin{abstract}
\baselineskip=0.6 cm
\begin{center}
{\bf Abstract}
\end{center}

We investigate the radiative efficiency and jet power in the spacetime of a rotating black hole within the framework of loop quantum gravity (LQG), which includes an additional LQG parameter. The results show that as the LQG parameter increases, the radiative efficiency decreases for slowly rotating black holes while it increases for rapidly rotating black holes. Furthermore, the jet power is found to increase for different black hole spins.
With the observed data from the well-known sources A0620-00, H1743-322, XTE J1550-564, GRS1124-683, GRO J1655-40, and GRS1915+105, we make some constraints on the black hole spin parameter and the LQG parameter. The presence of the LQG parameter  broadens the allowed range of the black hole spin parameter for sources A0620-00, H1743-322, XTE J1550-564 and GRO J1655-40. However, for the source GRS 1915+105, there is no overlap between the allowed parameter regions, which implies that the rotating LQG black hole cannot simultaneously account for the observed jet power and the radiative efficiency as in other black hole spacetimes.

\end{abstract}

\maketitle
\newpage
\section{Introduction}

It is believed widely that general relativity could be replaced by quantum gravity at the Planck scale, even if general relativity is the most successful theory to describe gravity at present. One of interesting quantum theories of gravity is loop quantum gravity (LQG) \cite{loop1,loop2,loop3} that preserves many
features of general relativity while simultaneously employing quantization of both space and time at the Planck scale in the tradition of quantum mechanics. The effects originating from LQG could resolve the singularity problem appeared in cosmology and black hole physics \cite{Adler:2010wf,Borde:1993xh,Emmanuele}.
Recently, based on the polymerization procedure in LQG, a regular and static black hole solution without
any spacetime curvature singularity has been derived through the mini-superspace approach \cite{Modesto:2008im}.  The effects of LQG in this solution are parameterized by the minimal area and the Barbero-Immirzi parameter. Additionally, the metric is self-dual in the context of T-duality because its form  is
invariant under the exchange $r\rightarrow a_0/r$ \cite{Modesto:2009ve}, where $a_0$ is proportional to the minimum area $A_{\rm min}$ in LQG and
$r$ is the standard Schwarzschild radial coordinate. The gravitational lensing \cite{Sahu:2015dea} and the emission spectra \cite{Hossenfelder} in this static self-dual black hole spacetime have been  investigated. The effects of LQG on quasinormal modes have been researched for various perturbation fields, including scalar field \cite{ChenJH,Santos2021,Momennia}, axial gravitational perturbations \cite{Cruz0,Yang:2023gas} and polar \cite{Cruz1} gravitational perturbations.
Observational tests of the self-dual spacetime have been performed within the frame of the Solar system \cite{Wangzhu1} as well as through the analysis of
the orbit of S0-2 star orbiting Sgr A* in the central region of our Milky Way \cite{Wangzhu2}. Since the real celestial bodies are inherently rotating, an effective rotating loop quantum black hole (LQBH) solution \cite{Liu:2020ola} has been obtained from the spherical symmetric LQBH using the modified Newman-Janis algorithm \cite{Azreg1,Azreg2}. The observable effects of LQG parameters, including black hole shadows and quasinormal modes, have been analyzed for this rotating spacetime \cite{Liu:2020ola}. Additionally, constraints on the parameters of this rotating LQG black hole have been studied using the observed data of Sgr A* from the Event Horizon Telescope \cite{Afrin1}.

Energetic transient jets are commonly observed in astrophysical systems containing black holes \cite{Mirabel,Fender}. While numerous theoretical models have been proposed to explain these jets, the complete mechanism is still absent at present. The Blandford-Znajeck mechanism is certainly a popular scenario  for explaining the formation of black hole jets. In this mechanism, black holes jets are assumed to be powered by the rotational energy of the black hole through the magnetic field whose field lines anchored in the horizon of black hole or in the accretion disk can be dragged and twisted
by the spin of the black hole \cite{ref69,ref77,Konoplya2021qll}. This means that the jet power could carry the information of the background black hole spacetime.  Moreover,
the continuum spectra released by the accretion disk near black holes contain the signals of the central celestial bodies \cite{ref70,ref78,PhysRevD86123013}.
And then, a novel method was firstly proposed \cite{PhysRevD86123013,ref84} to test the metric of black holes by combing the estimates of the jet powers with the estimates of the radiative efficiency of the disks. Utilizing this method along with relevant observational data, one can constrain the metric parameters and further evaluate corresponding theories of gravity. Based on the relation between jet power and spin  identified  in \cite{ref84,ref91,McClintock:2013vwa},  parameter constraints on the Kerr-Sen black hole were analyzed in \cite{add2}, suggesting  that  Kerr black holes are favored over Kerr-Sen black holes with dilaton charges. Additionally, using data on radiative efficiencies and jet powers, the suitability of the Kerr-Taub-NUT spacetime to describe the geometry around known sources \cite{ref98,ref100,steiner2011spin, ref109,Shafee06, ref118} was examined \cite{BakhtiyorNarzilloev}. Similar analyses were extended to the case of a rotating black hole within a perfect fluid dark matter model \cite{add1}. In this paper, we will investigate effects of the LQG parameter on the radiative efficiency and the power of relativistic jets from the Blandford-Znajeck mechanism in the rotating LQG black hole spacetime, and further probe  the possibility of indirect observational evidence of LQG by combining with the observed data from the known sources \cite{ref98,ref100,steiner2011spin, ref109,Shafee06, ref118}.

The paper is organized as follows:  In Sect.~\ref{section4.2}, we briefly review the rotating LQG black hole and probe effects of LQG parameter on the radiative efficiency and the power of relativistic jets in this background spacetime. In Sect.~\ref{section4.4}, we compare  the theoretical model with observations from the known sources \cite{ref98,ref100,steiner2011spin, ref109,Shafee06, ref118} and constrain the parameters of the black hole.
Finally, we present a summary.

\section{Effects of the LQG Parameter on Radiative Efficiency and Power of Relativistic Jets around a Rotating Black Hole\label{section4.2}}

Starting from the effective LQG-corrected Schwarzschild metric \cite{Modesto:2008im}, a rotating  black hole in LQG has been obtained by using the modified Newman-Janis algorithm \cite{Azreg1,Azreg2}. In the Boyer-Lindquist coordinates, its metric  can be expressed as \cite{Liu:2020ola},
\bqn
ds^2&=\frac{\mathscr{H}}{\Sigma} \bigg\{\frac{\Delta}{\Sigma}\bigg(dt-a\sin^2\theta d\phi\bigg)^2-\frac{\Sigma}{\Delta}dr^2-\Sigma d\theta^2-\frac{\sin^2\theta}{\Sigma}\bigg[a dt-(k^2+a^2)d\phi\bigg]^2 \bigg\},  \lb{mmm} \label{metric}
\eqn
with
\begin{eqnarray}
\Delta(r)&=& r^2\bigg[1-\frac{2GMr}{(r+r_*)^2}\bigg]+a^2, \quad\quad
\Sigma(r)=k^2(r)+a^2\cos^2\theta,\quad\quad
k^2(r)= \frac{r^4+a^2_0}{(r+r_*)^2},
\end{eqnarray}
where $M$ and $a$  denote  the mass and the spin parameter of the black hole, respectively. The quantity $r_{*}$ is related to the LQP parameter $P$ by the equation $r_{*}=2GMP/(1+P)^2$, where $P$ is defined as
\bqn
P \equiv \frac{\sqrt{1+\epsilon^2}-1}{\sqrt{1+\epsilon^2}+1}.   \label{P_epsilon}
\eqn
The quantity $\epsilon$ is defined as the product of the polymeric parameter $\delta$ and the Immirzi parameter $\gamma$, expressed as $\epsilon=\gamma \delta$. Its value is strictly constrained within the range $\epsilon=\gamma \delta \ll 1$, which ensures the validity of the effective metric derived from the polymerization procedure in LQG \cite{Modesto:2008im}.
 This means that the possible physical range of $P$ is given by $0\leq P\leq (\sqrt{2}-1)^2\approx 0.1716$.
The parameter $a_{0}$ is related to the minimum area gap of LQG, denoted as $A_{\rm min}$, by the equation $a_0 = A_{\rm min}/8\pi$, which implies that $a_0$ is proportional to $l_{Pl}$ and is expected to be negligible. Thus, we here consider only the case $a_0=0$.
The quantity $\mathscr{H}$  is a undefined  regular function in the metric (\ref{metric}), which must satisfy $\lim\limits_{a \to 0} \mathscr{H}=r^2+a_0^2/r^2$ \cite{Wangzhu1} and $\lim\limits_{P \to 0} \mathscr{H}=r^2+a^2\cos^2\theta$
because the  rotating LQG black hole (\ref{metric}) is expected to recover to the non-rotating self-dual black hole \cite{Wangzhu1} as $a=0$ and to the Kerr black hole as $P=0$. Combining with the previous discussion that $a_0$ can be negligible, we can set $a_0=0$ without loss of generality. Thus, we take the form
$\mathscr{H}=r^2+a^2\cos^2\theta$ as presented in \cite{Liu:2020ola}. In this case,
the event horizon radius of rotating LQG black hole (\ref{metric}) can be obtained by solving a simplified equation
\bqn
 (1 + P)^4 r^4-2M(1 + P)^2 (1 + P^2)r^3+[4 M^2 P^2 + a^2 (1 + P)^4] r^2+4 a^2 M P (1 + P)^2 r+4 a^2 M^2 P^2=0,
\eqn
which implies that the event horizon radius depends on the
values of $M$, $a$, and $P$. The explicit expressions of the outer  and inner horizon radii $r_{\pm}$ are given in \cite{Liu:2020ola}, which explore that
 the presence of the parameter $P$ results in a decrease in the value of $r_+$, while increasing the value of $r_-$. Thus, the dependence of horizon radii $r_{\pm}$ on the parameter $P$ is analogous to their dependence on the spin parameter $a$.
\begin{figure}[ht]
	\begin{center}
		\includegraphics[width=6cm]{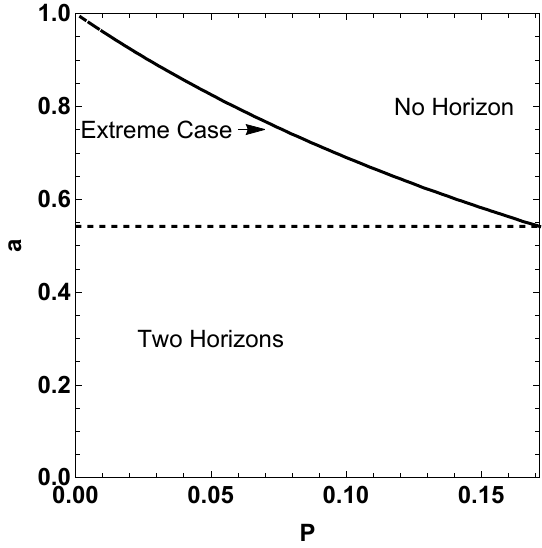}
	\end{center}
	\caption{Regions of existing two event horizons, one degenerate horizon  and no horizon in the physical parameter space of $P$ and $a$. The dashed line corresponds to the case $a=0.5413$. }\label{fww}
\end{figure}
In Fig.(\ref{fww}), we present the physical parameter space of $P$ and $a$, highlighting regions corresponding to two event horizons, one degenerate horizon, and no horizon. It is shown that as $0\leq a<0.5413$, the black hole always has two horizons and it is non-extremal in the physical region of $P\in [0, (\sqrt{2}-1)^2]$. As $a\geq0.5413$, the range of $P$ existing black hole solution decreases with the spin parameter $a$, which is consistent with that obtained in \cite{Liu:2020ola}. In the high spin case with $a=1.0$, there exists only an extremal black hole with $P=0$.

\begin{figure}[ht]
	\begin{center}
		\includegraphics[width=7cm]{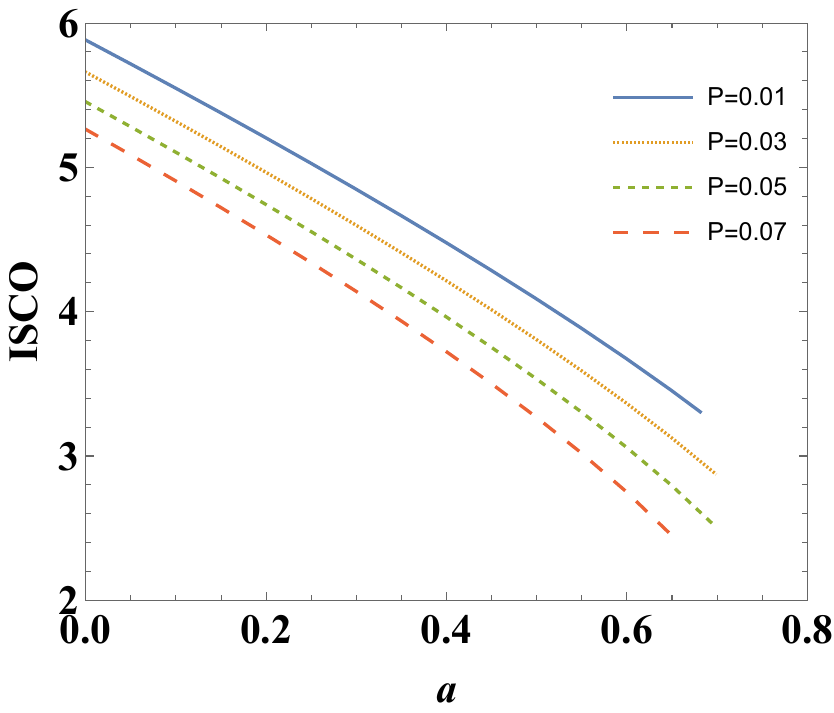}\quad\quad\includegraphics[width=7cm]{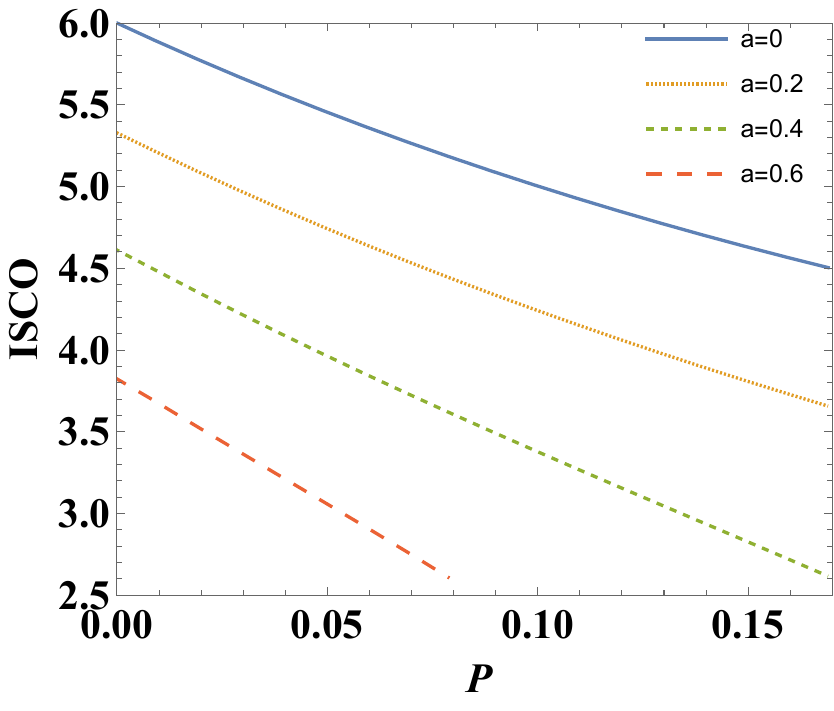}
	\end{center}
	\caption{Dependence of the ISCO radius on the parameters $a$ and $P$ of the rotating  black hole in LQG spacetime.}\label{fisco}
\end{figure}

Generally, the accretion process occurred near black holes are very complex. Here, we focus on only a simple Novikov-Thorne accretion disk model~\cite{ref70} .
The Novikov-Thorne accretion disk is geometrically thin, allowing the heat produced by viscous stresses and dynamic friction to be efficiently dissipated through radiation across its surface. This cooling mechanism ensures that the disk  remains in hydrodynamical equilibrium. In this geometrically thin disk model, 
 it is assumed that gravitational forces dominate gas motion over gas pressure, causing the particles in the disk to follow circular geodesic orbits in the equatorial plane.
The Novikov-Thorne disk exhibits a black body spectrum \cite{ref70,ref79,Bambi17e, 2021SSRv..217...65B}, which is highly sensitive to the position of the inner edge of the accretion disk. Generally, it is assumed that the inner edge of this thin disk is located at the Innermost Stable Circular Orbit (ISCO) radius.  With  the observed mass of the black hole, the inclination angle of the disk, and the distance between the black hole and the observer, one can infer the corresponding location of the ISCO radius \cite{Zhang_1997}.
 Theoretically, the ISCO radius can be inferred from the effective potential of a timelike particle orbiting a black hole.
For an axially symmetric and stationary spacetime (\ref{metric}), with the time-like condition of the particle 4-velocity $u_\mu u^\mu = -1$, one can obtain
\begin{eqnarray}
g_{rr} u_r^2 + g_{\theta\theta} u_\theta^2 = V_{\rm eff} ,
\end{eqnarray}
with the effective potential $V_{\rm eff}$
\begin{eqnarray}
V_{\rm eff}=\frac{E^2 g_{\phi \phi}+2 E L g_{t \phi}+L^2 g_{tt}}{g_{t\phi}^2-g_{tt}g_{\phi \phi}}-1,
\end{eqnarray}
 where $E$ and $L$ are respectively the conserved energy and the angular momentum of the orbiting particle. In term of the conditions for particles moving along circular orbits $V_{\rm eff}(r)=0$ and $V_{\rm eff}'(r)=0$ ( where the apostrophe $\prime$ denotes a derivative with respect to the radial coordinate $r$ ), one can obtain
\begin{eqnarray}
E&=&\frac{-g_{tt}-\Omega g_{t\phi}}{\sqrt{-g_{tt}-2\Omega g_{t\phi}-\Omega^2 g_{\phi \phi}}}\ ,\\
L&=&\frac{\Omega g_{\phi \phi}+g_{t \phi}}{\sqrt{-g_{tt}-2\Omega g_{t\phi}-\Omega^2 g_{\phi \phi}}}\ .
\end{eqnarray}
Here $\Omega = d\phi/dt$ is the angular velocity of the particle orbiting around the black hole~\cite{Bambi17e}
\beqn
\Omega=\frac{d\phi}{dt} = \frac{-g_{t\phi,r}\pm\sqrt{\{-g_{t\phi,r}\}^2-\{g_{\phi \phi, r}\} \{g_{tt,r}\}}}{g_{\phi \phi, r}}\
\eeqn
Together with the condition $V_{\rm eff}''(r)=0$, we can calculate the ISCO radius in the spacetime of a rotating  black hole in LQG (\ref{metric}).
Fig. (\ref{fisco})  illustrates the dependence of the ISCO radius on the parameters
$a$ and $P$ of the rotating  black hole in LQG spacetime,  indicating that the ISCO radius decreases as both black hole parameters increase.

\begin{figure}[ht]
	\begin{center}
		\includegraphics[width=7cm]{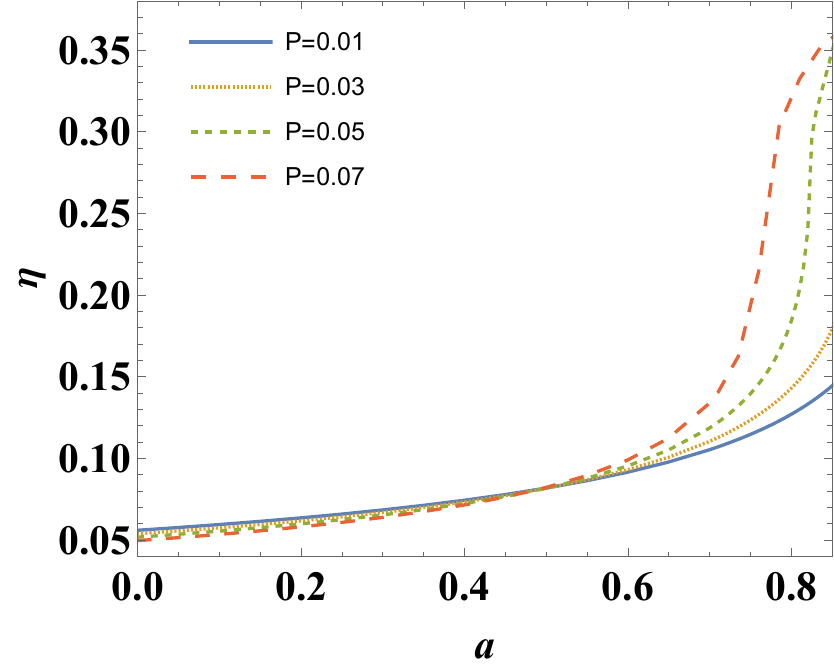}
	\end{center}
	\caption{Dependence of the conversion efficiency $\eta$  on the parameters $a$ and $P$ of the rotating  black hole in LQG spacetime.}\label{fetaa}
\end{figure}

In the Novikov-Thorne accretion disk,  all emitted photons can escape from the disk's surface to infinity.  Consequently, the radiative efficiency
$\eta$ is determined by the specific energy of a particle at the marginally stable orbit $r_{isco}$, i.e,
\begin{eqnarray}
\eta=1-E_{isco} , \label{etais}
\end{eqnarray}
The dependence of the conversion efficiency $\eta$ on the
parameters $a$ and $P$ is illustrated  in Fig. (\ref{fetaa}), which shows that
the conversion efficiency $\eta$ decreases with the LQG parameter $P$ for slowly rotating black holes,  while it increases for rapidly rotating black holes. Additionally, as the black hole spin increases, the conversion efficiency $\eta$ increases for different values of $P$, which is similar to that in the Kerr case.
It is well known that black holes surrounded by the Novikov-Thorne accretion disk with the same radiative efficiency must own the same thermal spectrum \cite{Kong14}, which can be utilized to estimate the parameters of the background black hole. Generally, to constrain the  parameters $a$ and $P$ in the black hole metric (\ref{metric}), we must firstly construct a theoretical thermal spectrum model of the disk surrounding the black hole, and compute the statistical quantity $\chi^2$ to find the best-fit values by comparing the observed spectra of the sources with the theoretical spectrum derived from the standard accretion disk model. This fitting method requires six free parameters: the black hole mass $M$ and its spin $a$, the LQG parameter $P$, the mass accretion rate $\dot{M}$, the black hole distance $d$, and the inclination angle $i$ of the disk with respect to the line of sight of the distant observer,
 which implies that this procedure is probably long and time-consuming \cite{Zhang_1997}. A simply approximate way to constraining black hole parameters is to estimate the radiative efficiency \cite{ref78,PhysRevD86123013}. The main reason is that the radiative efficiency depends on the spacetime metric and it directly connects the measured luminosity to the key physical parameter governing accretion dynamics. This method of applying radiative efficiency is extensively employed to constrain the black hole parameters in various scenarios \cite{ref78,PhysRevD86123013,ref84,ref91,McClintock:2013vwa,add2,BakhtiyorNarzilloev,add1}.

 Black hole jets are one of the most spectacular astronomical sights in the sky. Generally, jets near the central compact body are classified as the steady non-relativistic jets and the transiently relativistic jets. The latter are believed to originate from close to the event horizon \cite{ref77}, making them valuable tools for extracting information about the central black hole. However, the mechanisms that generate relativistic transient jets are so complicated that they are not fully understood at present ~\cite{ref85, ref87}. The Blandford-Znajeck
process is such a mechanism in which the relativistic jets are powered through extracting energy from the magnetic fields
around the accretion disk, which are dragged and twisted by the spin of the black hole. Here we employ the Blandford-Znajeck mechanism to estimate the power of transiently relativistic jets around the rotating LQG black hole (\ref{metric}). Assuming that the electromagnetic field dominates and other contributions can be neglected, the total energy-momentum tensor can be approximated as
\beqn
T_{\mu \nu}^{tot} \simeq T_{\mu \nu}^{EM}=F_{\mu \alpha} F^\alpha_\nu-\frac{1}{4}g_{\mu \nu} F_{\alpha \beta} F^{\alpha \beta},
\eeqn
where $F_{\mu\nu} = A_{\nu,\mu} - A_{\mu,\nu}$ is the electromagnetic field tensor related to the four potential $A_{\mu}$.
The corresponding covariant conservation equation can be simplified as
\beqn
\nabla^\mu T_{\mu \nu}^{EM}=0.
\eeqn
For a force-free magnetosphere, the electromagnetic  tensor satisfies
\beqn
F_{\mu\nu}J^{\nu}=0,\label{ffreeliu}
\eeqn
where $J^{\nu}$ is the current 4-vector. From Eq.(\ref{ffreeliu}), one can get the following relationship
\beqn
\frac{A_{t,r}}{A_\phi,r}=\frac{A_{t,\theta}}{A_{\phi,\theta}}=-\omega(r,\theta),\label{ffree}
\eeqn
where $\omega(r,\theta)$ is defined as the electromagnetic angular velocity~\cite{ref69}. With the force-free condition (\ref{ffree}), and the assumption that the four potential $A_{\mu}$ is stationary and axisymmetric, one can find that the electromagnetic  tensor  can be further expressed as
\beqn
F_{\mu \nu}=\sqrt{-g} \left(
\begin{array}{cccc}
 0 & - \omega B^{\theta }   &\omega B^r   & 0 \\
 \omega B^{\theta }  & 0 & B^{\phi } & -B^{\theta } \\
 -\omega B^r  & -B^{\phi } & 0 & B^r \\
 0 & B^{\theta } & -B^r & 0 \\
\end{array}
\right)\ .
\eeqn
In the context of the Blandford-Znajeck model, the power of the relativistic jets can be given by ~\cite{ref69}
\beqn
P_{BZ}=4 \pi \int_0^{\pi/2} \sqrt{-g} T_t^r d\theta \ ,
\eeqn
where $T_t^r$ is the radial component of the Poynting flux
\beqn
T_t^r=g^{rr}g^{\theta\theta}F_{r\theta}F_{\theta t}-g^{rt}g^{\theta\theta}F^2_{t\theta}+g^{r\phi}g^{\theta\theta}F_{\phi\theta}F_{\theta t}.
\eeqn
Thus, the spacetime affects the power of the relativistic jets through its metric determinant and the radial Poynting flux $T_t^r$.
 Assuming that the jet launching radius corresponds to the event horizon, the radial component of the Poynting flux can be further expressed as
\beqn
T_t^r=2 r_HM \sin^2\theta (B^r)^2 \omega [\Omega_H-\omega]|_{r=r_H} ,
\eeqn
where $\Omega_H$ is the angular velocity  evaluated at the event horizon $r_H$ with a form
\beqn
\Omega_H=-\frac{g_{t\phi}}{g_{\phi \phi}}\bigg|_{r_H}=\frac{ a \left(r_*+r_H\right)^2}{a^2 \left(r_*+r_H\right)^2+r_H^4}\, .
\eeqn
\begin{figure}[ht]
	\begin{center}
		\includegraphics[width=7cm]{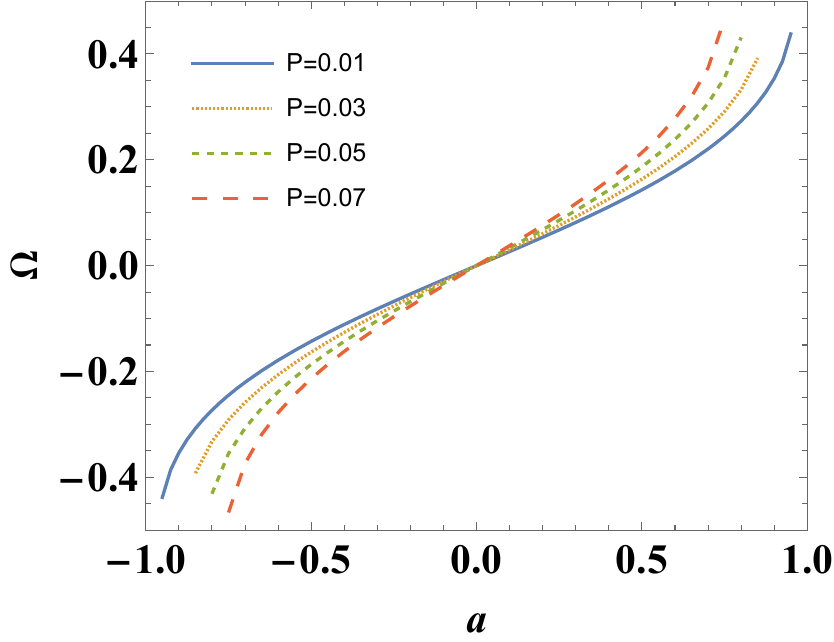}
	\end{center}
	\caption{Dependence of angular velocity at the event horizon $\Omega_H$  on the parameters $a$ and $P$ of the rotating  black hole in LQG spacetime.}\label{fomigaa}
\end{figure}
From Fig. (\ref{fomigaa}), we find that the absolute value of the angular velocity $\Omega_H$ at the event horizon increases with $P$.
For the Blandford-Znajek model, the leading-order contribution to the jet power~\cite{ref69,ref89,Camilloni:2022kmx}
is directly proportional to the square of  $\Omega_H$
\beqn
P_{BZ}=k \Phi_{tot}^2 \Omega_H^2\ , \label{PBZ}
\eeqn
where $k = 1/6\pi$ for a split monopole field profile and $k = 0.044$ for a paraboloidal one \cite{ref69}.  $\Phi_{tot}$ is the magnetic flux threading the event horizon, which is given by
\beqn
\Phi_{tot}=2 \pi \int_0^{\pi} \sqrt{-g} |B^r| d\theta\ .
\eeqn
Although this result is obtained in the Kerr metric \cite{ref89,Camilloni:2022kmx}, the analyses of the Blandford-Znajeck mechanism in other theories of gravity show that the specific gravity model affects only higher-order corrections, while the jet power at leading order retains the same form as given in Eq. (\ref{PBZ}) \cite{Camilloni:2023wyn}.
Since the jet power $P_{BZ}$ depends on the
metric through $\Omega_H$, we also present the change of angular velocity with the parameters $a$ and $P$ which characterize the gravitational field of the rotating black hole in loop quantum gravity (\ref{metric}).

\section{Comparison of theoretical models with observations \label{section4.4}}

The previous discussions show that both the radiative efficiency and the jet power  are sensitive to the spacetime metric. Therefore, through comparing theoretical models with the corresponding observations, one can obtain some signatures for rotating black holes in LQG and gain insights into the observationally favored range of the LQG parameter $P$. There are observational samples from six X-ray binaries including GRS1915+105,
GROJ1655-40, XTEJ1550-564, A0620-00, H1743-322 and GRS1124-683,  whose jet power and radiative efficiency are known from observations ~\cite{ref84,ref90,ref91}.
With these data,  the estimates of the spin parameter $a$ for the Kerr black hole and the corresponding radiative efficiency $\eta$ are listed in Table~\ref{Table1}.
\begin{widetext}
\begin{center}
\begin{table}[!ht]
\centering
    \begin{tabular}{|c|c|c|c|c|c|}
    \hline
    BH Source  & $a$ & $\eta$ 
    \\
    \hline
  A0620-00 & $0.12 { \pm 0.19}$~\cite{ref98} & $0.061^{+0.009}_{-0.007}$ 
  \\
    \hline
    H1743-322 & $0.2 { \pm 0.3}$ \cite{ref100}& $0.065^{+0.017}_{-0.011}$ 
    \\
    \hline
   XTEJ1550-564& $0.34 { \pm 0.24}$ \cite{steiner2011spin} & $0.072^{+0.017}_{-0.011}$ 
   \\
    \hline
   GRS1124-683 & $0.63^{+0.16}_{-0.19}$ \cite{ref109} & $0.095^{+0.025}_{-0.017}$ 
   \\
    \hline
   GROJ1655-40 & $0.7 { \pm 0.1}$ \cite{Shafee06}& $0.104^{+0.018}_{-0.013}$ 
   \\
    \hline
  GRS1915+105 & $ > 0.98$ \cite{ref118} & $ { > 0.234}$ 
  \\
    \hline
    \end{tabular}
  \caption{Parameters of the transient black hole binaries. The radiative efficiency $\eta$ is obtained from the spin measurement for the Kerr metric.}\label{Table1}
  \end{table}
  \end{center}
\end{widetext}

Follwing the procedure in~\cite{ref84,ref91},  one can calculate the jet power for the above six microquasars. Using the natural units, the proxy for the jet power is given by
\begin{eqnarray}
P_{jet}=\bigg(\frac{\nu}{5GHz}\bigg)\bigg(\frac{S^{tot}_{\nu,0}}{Jy}\bigg)\bigg(\frac{D}{kpc}\bigg)^2\bigg(\frac{M}{M_{\odot}}\bigg)^{-1},
\end{eqnarray}
 where $S^{tot}_{\nu,0}$ is the beaming corresponding to the approaching and receding jets. Here the entire power in the transient jet is assumed to be proportional to the peak 5 GHz radio flux density. To correct for the beaming, the Lorentz factor $\Gamma$ associated with the jet is anticipated to fall within the range $2\leq\Gamma \leq5$ ~\cite{Fender1,Fender2}, which commensurates with the mildly relativistic jets in microquasars. The Doppler corrected jet powers with the Lorenz factor $\Gamma=2$ and $\Gamma=5$ for the six black hole sources are respectively listed in Table \ref{Table2} ~\cite{ref90,ref94}.
\begin{table}[!ht]
    \large        
    \centering    
    \begin{tabular}{|c|c|c|c|}
    \hline
    BH Source & $P_{jet}|_{\Gamma=2}$ & $P_{jet}|_{\Gamma=5}$\\     
    \hline
    A0620-00 & 0.13 & 1.6\\
    \hline
    H1743-322 & 7.0 & 140\\
    \hline
     XTEJ1550-564 & 11 & 180\\
    \hline
     GRS1124-683 & 3.9 & 380\\
    \hline
     GROJ1655-40 & 70 & 1600\\
    \hline
     GRS1915+105 & 42 & 660\\
    \hline
    \end{tabular}
    \caption{Proxy jet power values in units of kpc$^2$~GHz~Jy~$M_{\odot}^{-1}$.}\label{Table2}
  \end{table}
From Eq.\eqref{PBZ}, the power of the jet can be rewritten as
\beqn\label{P}
\log P_{BZ}=\log K+2 \log \Omega_H ,
\eeqn
where $K = k \Phi_{tot}^2$. The magnitude of $K$ has been estimated by fitting above equation to the observed jet power $P_{BZ}$ as well as the angular velocity $\Omega_H$~\cite{ref84,ref94}. It is shown that $\log K = 2.94 \pm 0.22$ for the Lorentz factor $\Gamma=2$ and $\log K = 4.19 \pm 0.22$ for $\Gamma=5$ at $90\%$ confidence level ~\cite{ref94}.
Generally, the value of $K$ is  not a constant for all the sources. However, for the above six sources, each has a mass of approximately
 10 solar masses and the corresponding Eddington-scaled mass accretion rates are also similar because transient jets happen during the transition from the hard to the soft state.  This implies that for the above six sources the value of $K$ can be treated as a constant and is independent of the spacetime geometry. Therefore,
making use of these determined values of $K $ together with the observed jet
power of the sources in Table \ref{Table2}, we can make certain constraints on
the spin parameter $a$ and the LQG parameter $P$ of the rotating black hole (\ref{metric}).

\begin{figure}[ht]
	\begin{center}
	\includegraphics[width=6cm]{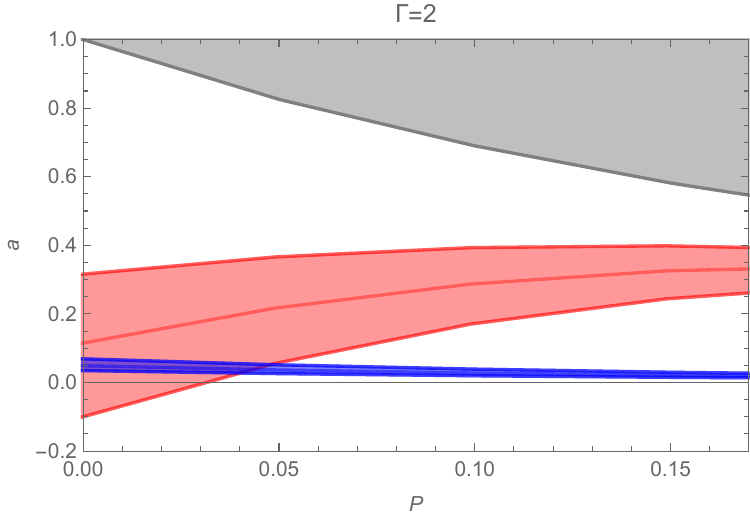}\quad\quad \includegraphics[width=6cm]{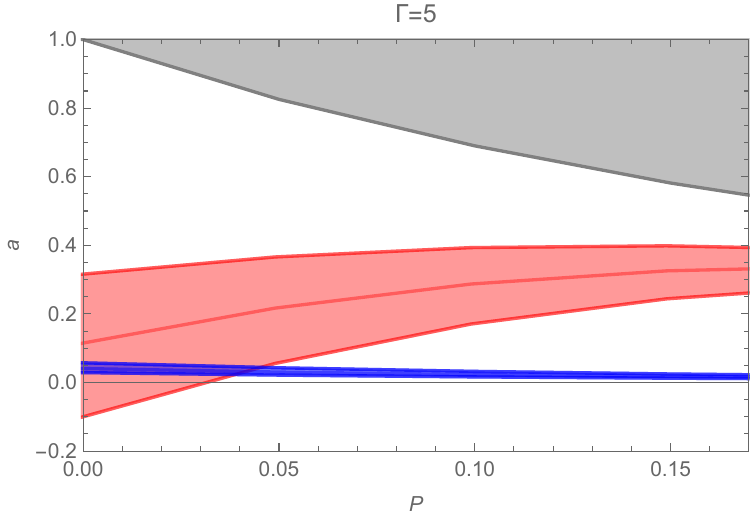}\\
    \includegraphics[width=6cm]{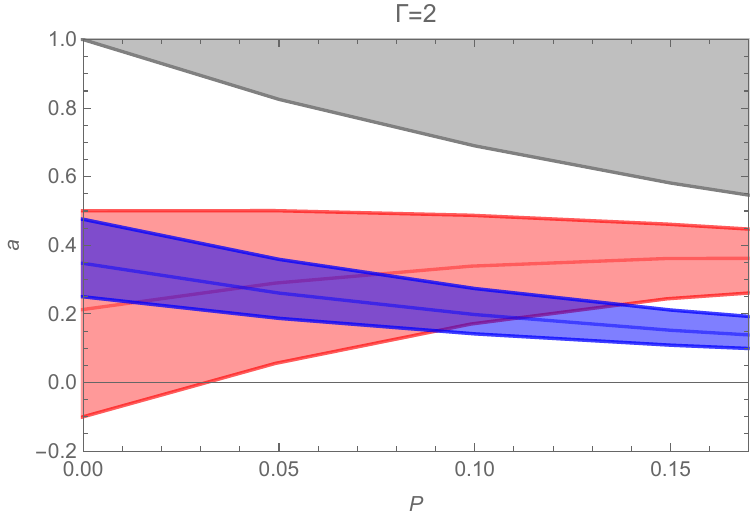}\quad\quad\includegraphics[width=6cm]{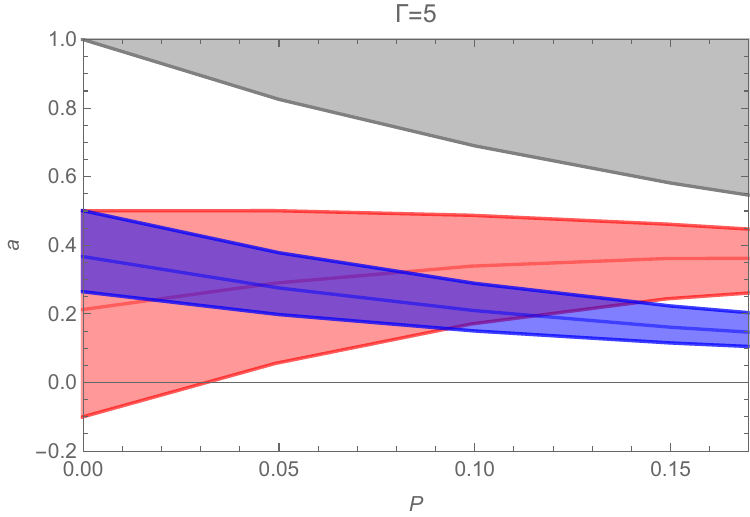}\\
    \includegraphics[width=6cm]{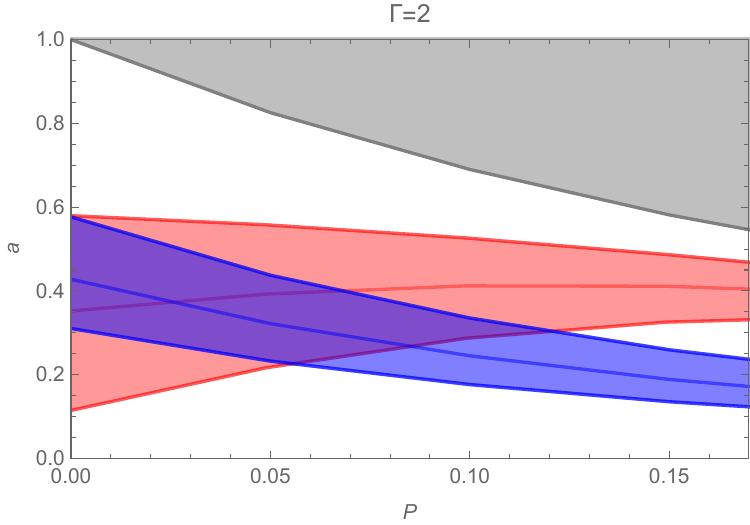}\quad\quad \includegraphics[width=6cm]{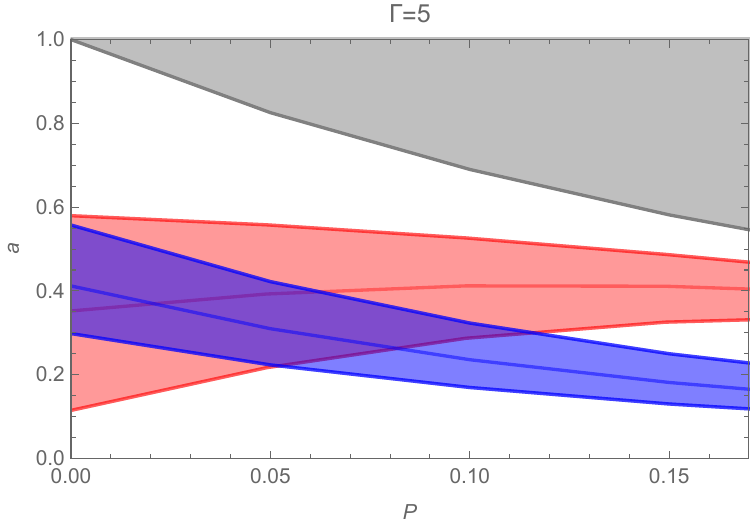}\\
    \includegraphics[width=6cm]{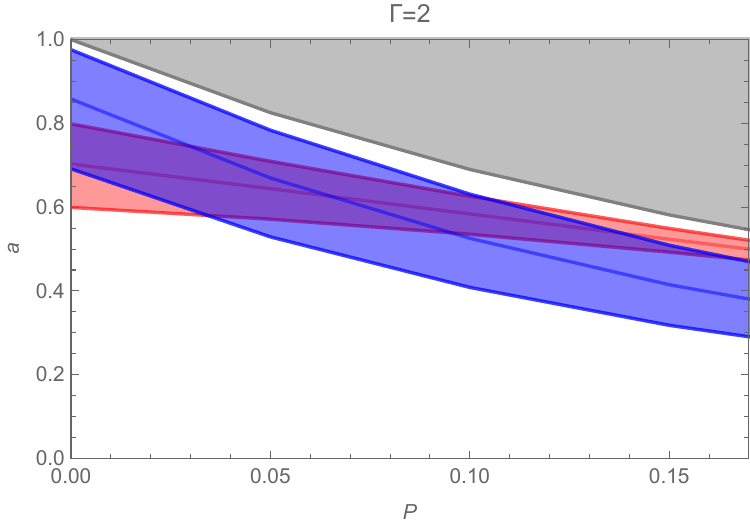}\quad\quad \includegraphics[width=6cm]{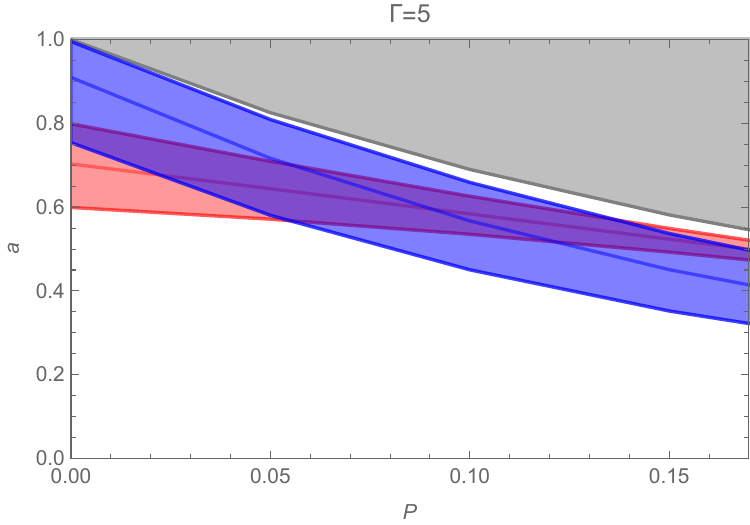}
	\end{center}
	\caption{The allowed regions in the parameter space $P-a$ for different sources. Panels from the top to the bottom respectively correspond to the sources A0620-00, H1743-322, XTE J1550-564 and GRO J1655-40. The red region and the blue region respectively correspond to the radiative efficiency and the jet power. Left panel is for the Lorentz factor $\Gamma=2$ and the right one for $\Gamma=5$.}\label{1Je}
\end{figure}

\begin{figure}[ht]
	\begin{center}
	\includegraphics[width=6cm]{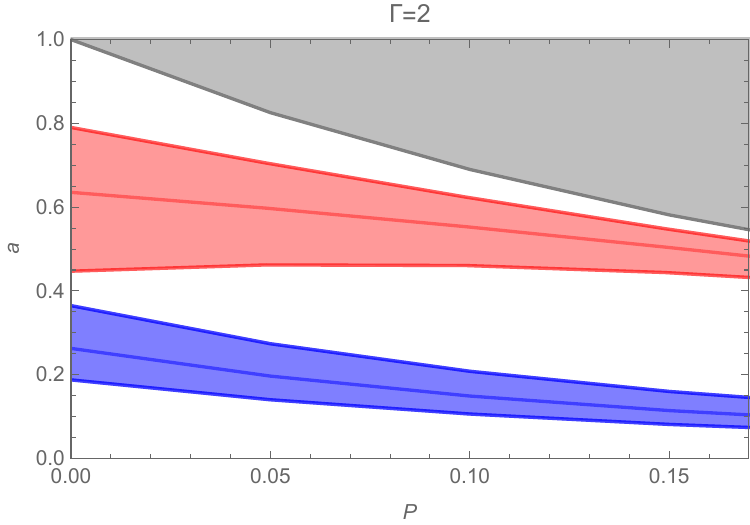}\quad\quad \includegraphics[width=6cm]{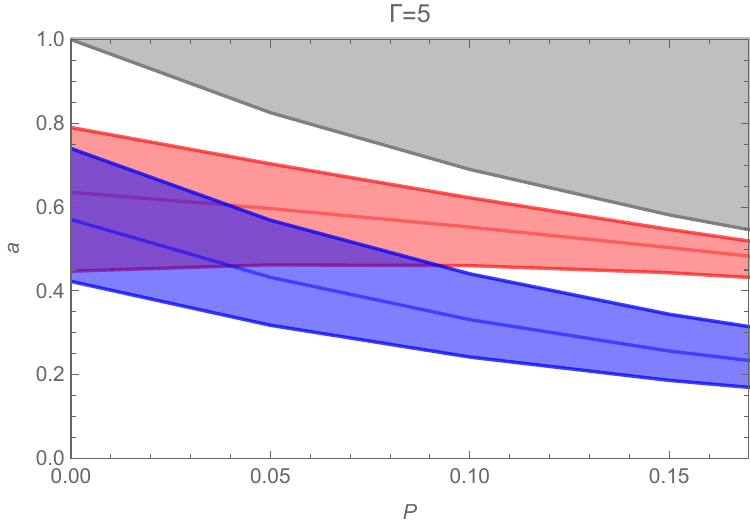}\\
    \includegraphics[width=6cm]{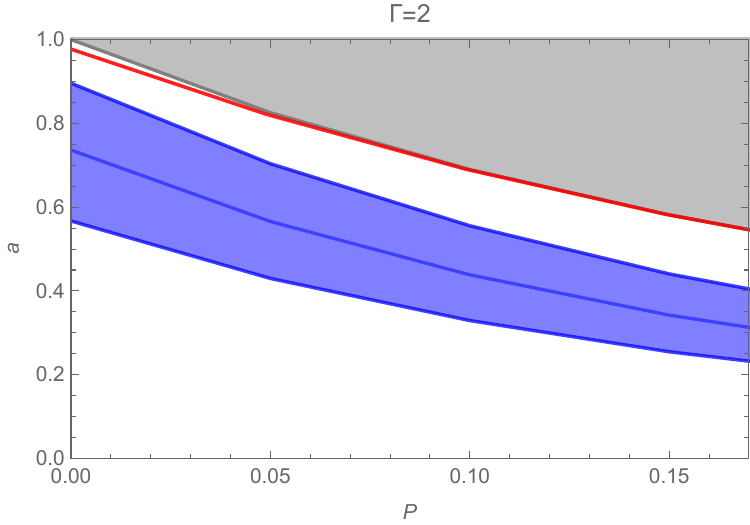}\quad\quad\includegraphics[width=6cm]{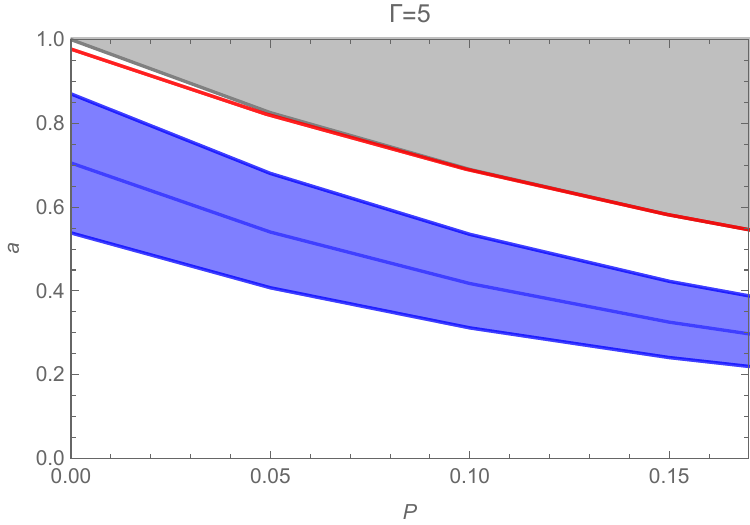}\\
	\end{center}
	\caption{The allowed area in the parameter space $P-a$ for the sources GRS 1124-683  and GRS1915+105. The top panels are for the source GRS 1124-683 and the bottom one are for GRS1915+105. The red region and the blue region respectively from the radiative efficiency and the jet power. Left panel is for the Lorentz factor $\Gamma=2$ and the right one for $\Gamma=5$.}\label{2Je}
\end{figure}

In Figs.(\ref{1Je}) and (\ref{2Je}), we present the allowed regions in the parameter space $P-a$ derived from the above six sources. In each panel, the red region and the blue region respectively correspond to the radiative efficiency and the jet power. The shade region indicates the scenario in which the compact object described by the rotating LQG metric (\ref{metric}) is a naked singularity rather than  a black hole.
With increasing of the LQG parameter $P$, the allowed value of the spin parameter $a$ for the source system A0620-00 increases for different radiative efficiency $\eta$. For the sources H1743-322 and XTE J1550-564, the allowed value of the spin parameter $a$ decreases for the cases with higher radiative efficiency $\eta$  and increases for the cases with lower efficiency. However, for the sources of GRO J1655-40, GRS 1124-683 and GRS~1915+105, the allowed value of the spin parameter $a$ decrease across all radiative efficiencies. The allowed value of $a$ from the jet power decreases with $P$ for all six sources. Moreover, the width of the allowed ranges of $a$  narrows with increasing $P$ from both the radiative efficiency and the jet power for these sources. The jet power provides the stronger constraint on $a$ for the source A0620-00, while the radiative efficiency offers a stronger constraint on $a$ for the source GRS~1915+105.

Combining with the two constrain regions from the radiative efficiency and the jet power, we find that there exist intersection regions in both cases the Lorentz factor $\Gamma=2$  and $\Gamma=5$ for the sources A0620-00, H1743-322, XTE J1550-564 and GRO J1655-40, which implies that the gravitational fields of these four sources could be described by the rotating LQG black hole (\ref{metric}). However, for the source GRS 1124-683, one can find the intersection of the constrain regions exists in the case $\Gamma=5$ but disappears in the case $\Gamma=2$. Especially, for the source GRS1915+105, there is no intersection for the two constrain regions from the radiative efficiency and the jet power in either cases $\Gamma=2$ or $\Gamma=5$. The results of the source GRS 1124-683 hint that the intersection regions could exist in the case with the more high Lorentz factor $\Gamma$ for  the source GRS1915+105.
The largest possible allowed ranges of the black hole parameters $a$ and $P$, derived from the intersection region between the radiative efficiency and the jet power, are listed in Table (\ref{table3}) for the six sources.
\begin{table}[ht]
\centering
    \begin{tabular}{|c|c|c|c|c|}
    \hline
      \multirow{2}{*}{BH Sources} &\multicolumn{2}{c}{$\Gamma=2$} & \multicolumn{2}{c|}{$\Gamma=5$} \\
    \cline{2-5}
    & $a$ & $P$ & $a$ & $P$
    \\
    \hline
    A0620-00 & $(0.03,\;0.07)$ & (0,\;0.048) & (0.02,\;0.06)& (0,\;0.046)
  \\
    \hline
    H1743-322 & (0.14,\;0.48)& (0,\;0.138) & (0.15,\;0.50)& (0,\;0.142)
    \\
    \hline
   XTEJ1550-564& (0.23,\;0.57)& (0.\;0.122) & (0.22,\;0.56) & (0.\;0.118)
   \\
    \hline
   GRS1124-683 & -- & -- & $(0.45,\;0.74) $&(0.\;0.092)
   \\
    \hline
   GROJ1655-40 & (0.57,\;0.8)& (0.\;0.166) &(0.56,\;0.70)& (0.\;0.171)
   \\
    \hline
  GRS1915+105 & -- & -- & --&--
  \\
    \hline
    \end{tabular}
  \caption{The largest possible allowed ranges of the black hole parameters $a$ and $P$ from the intersection region between the radiative efficiency and the jet power of the six different sources.}\label{table3}
  \end{table}
Comparing with the Kerr black hole, the presence of the LQG parameter $P$ broadens the allowed range of the black hole spin parameter $a$ for the sources  A0620-00, H1743-322, XTEJ1550-564 and  GROJ1655-40.   However, for the source GRS1124-683, one can find from the case $\Gamma=5$ that the largest allowed range of $a$ for the rotating LQG black hole (\ref{metric}) is the same as that in the Kerr case.  Moreover, the absence of an overlapping region corresponding to these two observational constraints for the source GRS1915+105 is also found in other spacetimes \cite{BakhtiyorNarzilloev,add1}. This leads to some arguments \cite{Russell2013ws} on the validity of the correlation between jet powers and black hole spins proposed by Narayan and McClintock in \cite{ref84} and \cite{ref91}. However, it is too early to assess  these arguments due to the limited number of available sources. With the increasing high precision observational data, it is expected that the validity of such kind of correlations could be confirmed in the future.

\section{Conclusion \label{section4.6}}

We have studied the radiative efficiency and the jet power in the spacetime of  the rotating LQG black hole (\ref{metric}). Our results show that with the increase of the LQG parameter $P$,  the conversion efficiency decreases for the slowly rotating black hole, but increases  for the rapidly rotating black hole.
The absolute value of the angular velocity $\Omega_H$ at the event horizon, associated with the jet power, increases with the LQG parameter $P$. With the observed data from the well-known sources A0620-00, H1743-322, XTE J1550-564, GRS1124-683, GRO J1655-40, and GRS1915+105, we also make some constraints on the black hole spin parameter $a$ and the LQG parameter $P$. It is found that the allowed ranges of $a$ decreases with $P$ based only on the radiative efficiency or from the jet power for these sources. The jet power provides the stronger constraint on $a$ for the source A0620-00, while  the radiative efficiency provides the stronger constraint on $a$ for the source GRS~1915+105.

With the data from both the radiative efficiency and the jet power, we find that there exist intersection regions in both cases the Lorentz factor $\Gamma=2$  and $\Gamma=5$ for the sources A0620-00, H1743-322, XTE J1550-564 and GRO J1655-40, which implies that the gravity of these four sources could be described by the rotating LQG black hole (\ref{metric}). The presence of the LQG parameter $P$ broadens the allowed range of the black hole spin parameter $a$ for these four sources. However, for the source GRS 1124-683, one can find the intersection of the constrain regions exists in the case $\Gamma=5$ but disappears in the case $\Gamma=2$. Especially, for the source GRS1915+105, there is no intersection for the two constrain regions respectively derived from the radiative efficiency and the jet power in both cases $\Gamma=2$ and $\Gamma=5$. The phenomenon of no overlapped region corresponding to these two observational constraints for the source GRS1915+105
is also found in other spacetimes \cite{BakhtiyorNarzilloev,add1}, which yields some arguments \cite{Russell2013ws} on the validity of the correlation between jet power and the black hole spin proposed by Narayan and McClintock in \cite{ref84} and \cite{ref91}. It is anticipated  that the validity of such kind of correlations could be confirmed in the future with the increasing high precision observational data.

\begin{acknowledgments}
	
This work was supported by the National Natural Science Foundation of China under Grant No.12275078, 11875026, 12035005, 2020YFC2201400, and the innovative research group of Hunan Province under Grant No. 2024JJ1006.

\end{acknowledgments}

\bibliographystyle{apsrev4-1}

\begin{thebibliography}{59}%
\makeatletter
\providecommand \@ifxundefined [1]{%
 \@ifx{#1\undefined}
}%
\providecommand \@ifnum [1]{%
 \ifnum #1\expandafter \@firstoftwo
 \else \expandafter \@secondoftwo
 \fi
}%
\providecommand \@ifx [1]{%
 \ifx #1\expandafter \@firstoftwo
 \else \expandafter \@secondoftwo
 \fi
}%
\providecommand \natexlab [1]{#1}%
\providecommand \enquote  [1]{``#1''}%
\providecommand \bibnamefont  [1]{#1}%
\providecommand \bibfnamefont [1]{#1}%
\providecommand \citenamefont [1]{#1}%
\providecommand \href@noop [0]{\@secondoftwo}%
\providecommand \href [0]{\begingroup \@sanitize@url \@href}%
\providecommand \@href[1]{\@@startlink{#1}\@@href}%
\providecommand \@@href[1]{\endgroup#1\@@endlink}%
\providecommand \@sanitize@url [0]{\catcode `\\12\catcode `\$12\catcode
  `\&12\catcode `\#12\catcode `\^12\catcode `\_12\catcode `\%12\relax}%
\providecommand \@@startlink[1]{}%
\providecommand \@@endlink[0]{}%
\providecommand \url  [0]{\begingroup\@sanitize@url \@url }%
\providecommand \@url [1]{\endgroup\@href {#1}{\urlprefix }}%
\providecommand \urlprefix  [0]{URL }%
\providecommand \Eprint [0]{\href }%
\providecommand \doibase [0]{http://dx.doi.org/}%
\providecommand \selectlanguage [0]{\@gobble}%
\providecommand \bibinfo  [0]{\@secondoftwo}%
\providecommand \bibfield  [0]{\@secondoftwo}%
\providecommand \translation [1]{[#1]}%
\providecommand \BibitemOpen [0]{}%
\providecommand \bibitemStop [0]{}%
\providecommand \bibitemNoStop [0]{.\EOS\space}%
\providecommand \EOS [0]{\spacefactor3000\relax}%
\providecommand \BibitemShut  [1]{\csname bibitem#1\endcsname}%
\let\auto@bib@innerbib\@empty


\bibitem{loop1} C. Rovelli, Living Rev. Rel. {\bf1}, 1 (1998) 1.
\bibitem{loop2} T. Thiemann, Lect. Notes Phys. {\bf 631}, 41 (2003).
\bibitem{loop3} A. Ashtekar, Class. Quant. Grav. {\bf 21},  R53 (2004).

\bibitem{Adler:2010wf}
R.~J.~Adler,
Am. J. Phys. \textbf{78} (2010), 925-932
doi:10.1119/1.3439650
[arXiv:1001.1205 [gr-qc]].

\bibitem{Borde:1993xh}
A.~Borde and A.~Vilenkin,
Phys. Rev. Lett. \textbf{72} (1994), 3305-3309
doi:10.1103/PhysRevLett.72.3305
[arXiv:gr-qc/9312022 [gr-qc]].

\bibitem{Emmanuele}E. Battista, \textit{Quantum Schwarzschild geometry in effective field theory models of gravity}, Phys. Rev. D {\bf 109}, 026004 (2024).

\bibitem{Modesto:2008im}
L.~Modesto,
Int. J. Theor. Phys. \textbf{49}, 1649-1683 (2010)
doi:10.1007/s10773-010-0346-x
[arXiv:0811.2196 [gr-qc]].



\bibitem{Modesto:2009ve}
L.~Modesto and I.~Premont-Schwarz,
Phys. Rev. D \textbf{80}, 064041 (2009)
doi:10.1103/PhysRevD.80.064041
[arXiv:0905.3170 [hep-th]].



\bibitem{Sahu:2015dea}
S.~Sahu, K.~Lochan and D.~Narasimha,
Phys. Rev. D \textbf{91}, 063001 (2015)
doi:10.1103/PhysRevD.91.063001
[arXiv:1502.05619 [gr-qc]].

\bibitem{Hossenfelder} S. Hossenfelder, L. Modesto, and I. Premont-Schwarz, \textit{Emission spectra of self-dual black holes}, [arXiv:1202.0412 [gr-qc]].

\bibitem{ChenJH} J. H. Chen and Y. J. Wang, \textit{Complex frequencies of a massless scalar fiel in loop quantum black hole spacetime}, Chin.
Phys. B {\bf20}, 030401 (2011).
\bibitem{Santos2021} J. S. Santos, M. B. Cruz, and F. A. Brito, \textit{Quasinormal modes of a massive scalar field nonminimally coupled to gravity
in the spacetime of self-dual black hole}, Eur. Phys. J. C {\bf81}, 1082 (2021).
\bibitem{Momennia}M. Momennia, \textit{Quasinormal modes of self-dual black holes in loop quantum gravity}, Phys. Rev. D {\bf106}, 024052 (2022),
[arXiv:2204.03259 [gr-qc]].

\bibitem{Cruz0} M. B. Cruz, C. A. S. Silva, and F. A. Brito, \textit{Gravitational axial perturbations and quasinormal modes of loop quantum
black holes}, Eur. Phys. J. C {\bf79}, 157 (2019), [arXiv:1511.08263 [gr-qc]].
\bibitem{Yang:2023gas} S.~Yang, W.~D.~Guo, Q.~Tan and Y.~X.~Liu, ``Axial gravitational quasinormal modes of a self-dual black hole in loop quantum gravity,'' Phys. Rev. D \textbf{108}, no.2, 024055 (2023) doi:10.1103/PhysRevD.108.024055 [arXiv:2304.06895 [gr-qc]]. 
\bibitem{Cruz1} M. B. Cruz, F. A. Brito, and C. A. S. Silva, \textit{Polar gravitational perturbations and quasinormal modes of a loop quantum
gravity black hole}, Phys. Rev. D {\bf102}, 044063 (2020), [arXiv:2005.02208 [gr-qc]].

\bibitem{Wangzhu1}T. Zhu and A. Wang, \textit{Observational tests of the self-dual spacetime in loop quantum gravity}, Phys. Rev. D {\bf102}, 124042
(2020), [arXiv:2008.08704 [gr-qc]].
\bibitem{Wangzhu2} J. M. Yan, Q. Wu, C. Liu, T. Zhu, and A. Wang, \textit{Constraints on self-dual black hole in loop quantum gravity with S0-2
star in the galactic center}, JCAP {\bf09}, 008 (2022), [arXiv:2203.03203 [gr-qc]].

\bibitem{Liu:2020ola}
C.~Liu, T.~Zhu, Q.~Wu, K.~Jusufi, M.~Jamil, M.~Azreg-A\"{i}nou and A.~Wang,
\textit{Shadow and Quasinormal Modes of a Rotating Loop Quantum Black Hole},
Phys. Rev. D \textbf{101}, 084001 (2020).
	
\bibitem{Azreg1} M. Azreg-A\"{\i}nou, \textit{Generating rotating regular black hole solutions without complexification}, Phys. Rev. D {\bf90}, 064041 (2014).
\bibitem{Azreg2} M. Azreg-A\"{\i}nou, \textit{From static to rotating to conformal static solutions: rotating imperfect fluid wormholes with(out) electric or magnetic field}, Eur. Phys. J. C {\bf74}, 2865 (2014).

\bibitem{Afrin1} M. Afrin, S. Vagnozzi,  S. G. Ghosh\textit, {Tests of Loop Quantum Gravity from the Event Horizon Telescope Results of Sgr A*},  Astrophys. J. {\bf944},149 (2023).


\bibitem{Mirabel} I. Mirabel and L. Rodriguez, \textit{A Superluminal source in the galaxy}, Nature {\bf371},46 (1994).
\bibitem{Fender}R. Fender and T. Belloni, \textit{GRS 1915+105 and the disc-jet coupling in accreting black hole
systems}, Ann. Rev. Astron. Astrophys. {\bf42}, 317 (2004).  arXiv:astro-ph/0406483.

\bibitem [{\citenamefont {Blandford}\ and\ \citenamefont
  {Znajek}(1977)}]{ref69}%
  \BibitemOpen
  \bibfield  {author} {\bibinfo {author} {\bibfnamefont {R.~D.}\ \bibnamefont
  {Blandford}}\ and\ \bibinfo {author} {\bibfnamefont {R.~L.}\ \bibnamefont
  {Znajek}},\ }
  {{ {Mon. Not. Roy. Astron. Soc.}\ }\textbf { {179}},\  {433} ({1977})}%
\bibitem [{\citenamefont {Mirabel}\ and\ \citenamefont
  {Rodriguez}(1999)}]{ref77}%
  \BibitemOpen
  \bibfield  {author} {\bibinfo {author} {\bibfnamefont {I.~F.}\ \bibnamefont
  {Mirabel}}\ and\ \bibinfo {author} {\bibfnamefont {L.~F.}\ \bibnamefont
  {Rodriguez}},\ }
  {\bibfield
  {journal} {\bibinfo  {journal} {Ann. Rev. Astron. Astrophys.}\ }\textbf
  {\bibinfo {volume} {37}},\ \bibinfo {pages} {409} (\bibinfo {year} {1999})},\
   {arXiv:astro-ph/9902062}
  \BibitemShut {NoStop}%
\bibitem [{\citenamefont {Konoplya}\ \emph {et~al.}(2021)\citenamefont
  {Konoplya}, \citenamefont {Kunz},\ and\ \citenamefont
  {Zhidenko}}]{Konoplya2021qll}%
  \BibitemOpen
  \bibfield  {author} {\bibinfo {author} {\bibfnamefont {R.~A.}\ \bibnamefont
  {Konoplya}}, \bibinfo {author} {\bibfnamefont {J.}~\bibnamefont {Kunz}}, \
  and\ \bibinfo {author} {\bibfnamefont {A.}~\bibnamefont {Zhidenko}},\
  }\href@noop {} {\  (\bibinfo {year} {2021})},\
  {arXiv:2102.10649 [gr-qc]} \BibitemShut
  {NoStop}%
\bibitem [{\citenamefont {Novikov}\ and\ \citenamefont {Thorne}(1973)}]{ref70}%
  \BibitemOpen
  \bibfield  {author} {\bibinfo {author} {\bibfnamefont {I.~D.}\ \bibnamefont
  {Novikov}}\ and\ \bibinfo {author} {\bibfnamefont {K.~S.}\ \bibnamefont
  {Thorne}},\ }in\ \href@noop {} {\emph {\bibinfo {booktitle} {{Les Houches
  Summer School of Theoretical Physics}: {Black Holes}}}}\ (\bibinfo {year}
  {1973})\ pp.\ \bibinfo {pages} {343--550}\BibitemShut {NoStop}%
\bibitem [{\citenamefont {Bambi}(2012{\natexlab{a}})}]{ref78}%
  \BibitemOpen
  \bibfield  {author} {\bibinfo {author} {\bibfnamefont {C.}~\bibnamefont
  {Bambi}},\ }  {\bibfield  {journal}
  {\bibinfo  {journal} {Phys. Rev. D}\ }\textbf {\bibinfo {volume} {85}},\
  \bibinfo {pages} {043002} (\bibinfo {year} {2012}{\natexlab{a}})},\
   {arXiv:1201.1638 [gr-qc]} \BibitemShut
  {NoStop}%
\bibitem [{\citenamefont {Bambi}(2012{\natexlab{b}})}]{PhysRevD86123013}%
  \BibitemOpen
  \bibfield  {author} {\bibinfo {author} {\bibfnamefont {C.}~\bibnamefont
  {Bambi}},\ }  {\bibfield  {journal}
  {\bibinfo  {journal} {Phys. Rev. D}\ }\textbf {\bibinfo {volume} {86}},\
  \bibinfo {pages} {123013} (\bibinfo {year} {2012}{\natexlab{b}})}\BibitemShut
  {NoStop}%
\bibitem [{\citenamefont {Narayan}\ and\ \citenamefont
  {McClintock}(2012)}]{ref84}%
  \BibitemOpen
  \bibfield  {author} {\bibinfo {author} {\bibfnamefont {R.}~\bibnamefont
  {Narayan}}\ and\ \bibinfo {author} {\bibfnamefont {J.~E.}\ \bibnamefont
  {McClintock}},\ } {\bibfield
   {journal} {\bibinfo  {journal} {Mon. Not. Roy. Astron. Soc.}\ }\textbf
  {\bibinfo {volume} {419}},\ \bibinfo {pages} {L69} (\bibinfo {year}
  {2012})},\   {arXiv:1112.0569
  [astro-ph.HE]} \BibitemShut {NoStop}%
\bibitem [{\citenamefont {Steiner}\ \emph {et~al.}(2013)\citenamefont
  {Steiner}, \citenamefont {McClintock},\ and\ \citenamefont
  {Narayan}}]{ref91}%
  \BibitemOpen
  \bibfield  {author} {\bibinfo {author} {\bibfnamefont {J.~F.}\ \bibnamefont
  {Steiner}}, \bibinfo {author} {\bibfnamefont {J.~E.}\ \bibnamefont
  {McClintock}}, \ and\ \bibinfo {author} {\bibfnamefont {R.}~\bibnamefont
  {Narayan}},\ }{\bibfield
  {journal} {\bibinfo  {journal} {Astrophys. J.}\ }\textbf {\bibinfo {volume}
  {762}},\ \bibinfo {pages} {104} (\bibinfo {year} {2013})},\  {arXiv:1211.5379 [astro-ph.HE]} \BibitemShut
  {NoStop}%
\bibitem [{\citenamefont {McClintock}\ \emph {et~al.}(2014)\citenamefont
  {McClintock}, \citenamefont {Narayan},\ and\ \citenamefont
  {Steiner}}]{McClintock:2013vwa}%
  \BibitemOpen
  \bibfield  {author} {\bibinfo {author} {\bibfnamefont {J.~E.}\ \bibnamefont
  {McClintock}}, \bibinfo {author} {\bibfnamefont {R.}~\bibnamefont {Narayan}},
  \ and\ \bibinfo {author} {\bibfnamefont {J.~F.}\ \bibnamefont {Steiner}},\
  } {\bibfield  {journal} {\bibinfo
  {journal} {Space Sci. Rev.}\ }\textbf {\bibinfo {volume} {183}},\ \bibinfo
  {pages} {295} (\bibinfo {year} {2014})},\  {arXiv:1303.1583 [astro-ph.HE]} \BibitemShut
  {NoStop}%
\bibitem{add2}I. Banerjee, B. Mandal, S. SenGupta, \textit{Signatures of Einstein-Maxwell dilaton-axion gravity from the observed jet power and the radiative efficiency}, Phys. Rev. D {\bf103}, 044046 (2021).


\bibitem [{\citenamefont {Gou}\ \emph {et~al.}(2010)\citenamefont {Gou},
  \citenamefont {McClintock}, \citenamefont {Steiner}, \citenamefont {Narayan},
  \citenamefont {Cantrell}, \citenamefont {Bailyn},\ and\ \citenamefont
  {Orosz}}]{ref98}%
  \BibitemOpen
  \bibfield  {author} {\bibinfo {author} {\bibfnamefont {L.}~\bibnamefont
  {Gou}}, \bibinfo {author} {\bibfnamefont {J.~E.}\ \bibnamefont {McClintock}},
  \bibinfo {author} {\bibfnamefont {J.~F.}\ \bibnamefont {Steiner}}, \bibinfo
  {author} {\bibfnamefont {R.}~\bibnamefont {Narayan}}, \bibinfo {author}
  {\bibfnamefont {A.~G.}\ \bibnamefont {Cantrell}}, \bibinfo {author}
  {\bibfnamefont {C.~D.}\ \bibnamefont {Bailyn}}, \ and\ \bibinfo {author}
  {\bibfnamefont {J.~A.}\ \bibnamefont {Orosz}},\ } {\bibfield  {journal} {\bibinfo  {journal}
  {Astrophys. J. Lett.}\ }\textbf {\bibinfo {volume} {718}},\ \bibinfo {pages}
  {L122} (\bibinfo {year} {2010})},\
  {arXiv:1002.2211 [astro-ph.HE]} \BibitemShut {NoStop}%
\bibitem [{\citenamefont {Steiner}\ \emph {et~al.}(2012)\citenamefont
  {Steiner}, \citenamefont {McClintock},\ and\ \citenamefont {Reid}}]{ref100}%
  \BibitemOpen
  \bibfield  {author} {\bibinfo {author} {\bibfnamefont {J.~F.}\ \bibnamefont
  {Steiner}}, \bibinfo {author} {\bibfnamefont {J.~E.}\ \bibnamefont
  {McClintock}}, \ and\ \bibinfo {author} {\bibfnamefont {M.~J.}\ \bibnamefont
  {Reid}},\ } {\bibfield  {journal}
  {\bibinfo  {journal} {Astrophys. J. Lett.}\ }\textbf {\bibinfo {volume}
  {745}},\ \bibinfo {pages} {L7} (\bibinfo {year} {2012})},\  {arXiv:1111.2388 [astro-ph.HE]} \BibitemShut
  {NoStop}%
\bibitem [{\citenamefont {Steiner}\ \emph {et~al.}(2011)\citenamefont
  {Steiner}, \citenamefont {Reis}, \citenamefont {McClintock}, \citenamefont
  {Narayan}, \citenamefont {Remillard}, \citenamefont {Orosz}, \citenamefont
  {Gou}, \citenamefont {Fabian},\ and\ \citenamefont
  {Torres}}]{steiner2011spin}%
  \BibitemOpen
  \bibfield  {author} {\bibinfo {author} {\bibfnamefont {J.~F.}\ \bibnamefont
  {Steiner}}, \bibinfo {author} {\bibfnamefont {R.~C.}\ \bibnamefont {Reis}},
  \bibinfo {author} {\bibfnamefont {J.~E.}\ \bibnamefont {McClintock}},
  \bibinfo {author} {\bibfnamefont {R.}~\bibnamefont {Narayan}}, \bibinfo
  {author} {\bibfnamefont {R.~A.}\ \bibnamefont {Remillard}}, \bibinfo {author}
  {\bibfnamefont {J.~A.}\ \bibnamefont {Orosz}}, \bibinfo {author}
  {\bibfnamefont {L.}~\bibnamefont {Gou}}, \bibinfo {author} {\bibfnamefont
  {A.~C.}\ \bibnamefont {Fabian}}, \ and\ \bibinfo {author} {\bibfnamefont
  {M.~A.}\ \bibnamefont {Torres}},\ }\href@noop {} {\bibfield  {journal}
  {\bibinfo  {journal} {Monthly Notices of the Royal Astronomical Society}\
  }\textbf {\bibinfo {volume} {416}},\ \bibinfo {pages} {941} (\bibinfo {year}
  {2011})}\BibitemShut {NoStop}%
\bibitem [{\citenamefont {Chen}\ \emph {et~al.}(2016)\citenamefont {Chen},
  \citenamefont {Gou}, \citenamefont {McClintock}, \citenamefont {Steiner},
  \citenamefont {Wu}, \citenamefont {Xu}, \citenamefont {Orosz},\ and\
  \citenamefont {Xiang}}]{ref109}%
  \BibitemOpen
  \bibfield  {author} {\bibinfo {author} {\bibfnamefont {Z.}~\bibnamefont
  {Chen}}, \bibinfo {author} {\bibfnamefont {L.}~\bibnamefont {Gou}}, \bibinfo
  {author} {\bibfnamefont {J.~E.}\ \bibnamefont {McClintock}}, \bibinfo
  {author} {\bibfnamefont {J.~F.}\ \bibnamefont {Steiner}}, \bibinfo {author}
  {\bibfnamefont {J.}~\bibnamefont {Wu}}, \bibinfo {author} {\bibfnamefont
  {W.}~\bibnamefont {Xu}}, \bibinfo {author} {\bibfnamefont {J.}~\bibnamefont
  {Orosz}}, \ and\ \bibinfo {author} {\bibfnamefont {Y.}~\bibnamefont
  {Xiang}},\ } {\bibfield  {journal}
  {\bibinfo  {journal} {Astrophys. J.}\ }\textbf {\bibinfo {volume} {825}},\
  \bibinfo {pages} {45} (\bibinfo {year} {2016})},\  {arXiv:1601.00615 [astro-ph.HE]}
  \BibitemShut {NoStop}%
\bibitem [{\citenamefont {{Shafee}}\ \emph {et~al.}(2006)\citenamefont
  {{Shafee}}, \citenamefont {{McClintock}}, \citenamefont {{Narayan}},
  \citenamefont {{Davis}}, \citenamefont {{Li}},\ and\ \citenamefont
  {{Remillard}}}]{Shafee06}%
  \BibitemOpen
  \bibfield  {author} {\bibinfo {author} {\bibfnamefont {R.}~\bibnamefont
  {{Shafee}}}, \bibinfo {author} {\bibfnamefont {J.~E.}\ \bibnamefont
  {{McClintock}}}, \bibinfo {author} {\bibfnamefont {R.}~\bibnamefont
  {{Narayan}}}, \bibinfo {author} {\bibfnamefont {S.~W.}\ \bibnamefont
  {{Davis}}}, \bibinfo {author} {\bibfnamefont {L.-X.}\ \bibnamefont {{Li}}}, \
  and\ \bibinfo {author} {\bibfnamefont {R.~A.}\ \bibnamefont {{Remillard}}},\
  } {\bibfield  {journal} {\bibinfo  {journal}
  {Astrophys. J. Lett}\ }\textbf {\bibinfo {volume} {636}},\ \bibinfo {pages}
  {L113} (\bibinfo {year} {2006})},\  {astro-ph/0508302} \BibitemShut
  {NoStop}%
\bibitem [{\citenamefont {McClintock}\ \emph {et~al.}(2006)\citenamefont
  {McClintock}, \citenamefont {Shafee}, \citenamefont {Narayan}, \citenamefont
  {Remillard}, \citenamefont {Davis},\ and\ \citenamefont {Li}}]{ref118}%
  \BibitemOpen
  \bibfield  {author} {\bibinfo {author} {\bibfnamefont {J.~E.}\ \bibnamefont
  {McClintock}}, \bibinfo {author} {\bibfnamefont {R.}~\bibnamefont {Shafee}},
  \bibinfo {author} {\bibfnamefont {R.}~\bibnamefont {Narayan}}, \bibinfo
  {author} {\bibfnamefont {R.~A.}\ \bibnamefont {Remillard}}, \bibinfo {author}
  {\bibfnamefont {S.~W.}\ \bibnamefont {Davis}}, \ and\ \bibinfo {author}
  {\bibfnamefont {L.-X.}\ \bibnamefont {Li}},\ }
  {\bibfield  {journal} {\bibinfo  {journal} {Astrophys. J.}\ }\textbf
  {\bibinfo {volume} {652}},\ \bibinfo {pages} {518} (\bibinfo {year}
  {2006})},\
  {arXiv:astro-ph/0606076} \BibitemShut {NoStop}%
\bibitem{BakhtiyorNarzilloev}
B. Narzilloev, A. Abdujabbarov, B. Ahmedov, and C. Bambi, \textit{Kerr-Taub-NUT spacetime to explain the jet power and the radiative efficiency of astrophysical black holes}, Phys. Rev. D \textbf{108}, 103013 (2023).
\bibitem{add1} B. Narzilloev, A. Abdujabbarov, B. Ahmedov, C. Bambi,\textit{Observed jet power and radiative efficiency of black hole candidates in Kerr + PFDM model}, arXiv: 2408.05576.
\bibitem [{\citenamefont {Page}\ and\ \citenamefont {Thorne}(1974)}]{ref79}%
  \BibitemOpen
  \bibfield  {author} {\bibinfo {author} {\bibfnamefont {D.~N.}\ \bibnamefont
  {Page}}\ and\ \bibinfo {author} {\bibfnamefont {K.~S.}\ \bibnamefont
  {Thorne}},\ } {\bibfield  {journal} {\bibinfo
  {journal} {Astrophys. J.}\ }\textbf {\bibinfo {volume} {191}},\ \bibinfo
  {pages} {499} (\bibinfo {year} {1974})}\BibitemShut {NoStop}%
\bibitem [{\citenamefont {Bambi}(2017)}]{Bambi17e}%
  \BibitemOpen
  \bibfield  {author} {\bibinfo {author} {\bibfnamefont {C.}~\bibnamefont
  {Bambi}},\ }\href@noop {} {\emph {\bibinfo {title} {Black Holes: A Laboratory
  for Testing Strong Gravity}}}\ (\bibinfo  {publisher} {Springer, Singapore},\
  \bibinfo {year} {2017})\BibitemShut {NoStop}%
\bibitem [{\citenamefont {{Bambi}}\ \emph {et~al.}(2021)\citenamefont
  {{Bambi}}, \citenamefont {{Brenneman}}, \citenamefont {{Dauser}},
  \citenamefont {{Garc{\'\i}a}}, \citenamefont {{Grinberg}}, \citenamefont
  {{Ingram}}, \citenamefont {{Jiang}}, \citenamefont {{Liu}}, \citenamefont
  {{Lohfink}}, \citenamefont {{Marinucci}}, \citenamefont {{Mastroserio}},
  \citenamefont {{Middei}}, \citenamefont {{Nampalliwar}}, \citenamefont
  {{Nied{\'z}wiecki}}, \citenamefont {{Steiner}}, \citenamefont {{Tripathi}},\
  and\ \citenamefont {{Zdziarski}}}]{2021SSRv..217...65B}%
  \BibitemOpen
  \bibfield  {author} {\bibinfo {author} {\bibfnamefont {C.}~\bibnamefont
  {{Bambi}}}, \bibinfo {author} {\bibfnamefont {L.~W.}\ \bibnamefont
  {{Brenneman}}}, \bibinfo {author} {\bibfnamefont {T.}~\bibnamefont
  {{Dauser}}}, \bibinfo {author} {\bibfnamefont {J.~A.}\ \bibnamefont
  {{Garc{\'\i}a}}}, \bibinfo {author} {\bibfnamefont {V.}~\bibnamefont
  {{Grinberg}}}, \bibinfo {author} {\bibfnamefont {A.}~\bibnamefont
  {{Ingram}}}, \bibinfo {author} {\bibfnamefont {J.}~\bibnamefont {{Jiang}}},
  \bibinfo {author} {\bibfnamefont {H.}~\bibnamefont {{Liu}}}, \bibinfo
  {author} {\bibfnamefont {A.~M.}\ \bibnamefont {{Lohfink}}}, \bibinfo {author}
  {\bibfnamefont {A.}~\bibnamefont {{Marinucci}}}, \bibinfo {author}
  {\bibfnamefont {G.}~\bibnamefont {{Mastroserio}}}, \bibinfo {author}
  {\bibfnamefont {R.}~\bibnamefont {{Middei}}}, \bibinfo {author}
  {\bibfnamefont {S.}~\bibnamefont {{Nampalliwar}}}, \bibinfo {author}
  {\bibfnamefont {A.}~\bibnamefont {{Nied{\'z}wiecki}}}, \bibinfo {author}
  {\bibfnamefont {J.~F.}\ \bibnamefont {{Steiner}}}, \bibinfo {author}
  {\bibfnamefont {A.}~\bibnamefont {{Tripathi}}}, \ and\ \bibinfo {author}
  {\bibfnamefont {A.~A.}\ \bibnamefont {{Zdziarski}}},\ } {\bibfield  {journal} {\bibinfo  {journal} {Space
  Sci. Rev.}\ }\textbf {\bibinfo {volume} {217}},\ \bibinfo {eid} {65}
  (\bibinfo {year} {2021})},\
  {arXiv:2011.04792 [astro-ph.HE]} \BibitemShut {NoStop}%
\bibitem [{\citenamefont {Zhang}\ \emph {et~al.}(1997)\citenamefont {Zhang},
  \citenamefont {Cui},\ and\ \citenamefont {Chen}}]{Zhang_1997}%
  \BibitemOpen
  \bibfield  {author} {\bibinfo {author} {\bibfnamefont {S.~N.}\ \bibnamefont
  {Zhang}}, \bibinfo {author} {\bibfnamefont {W.}~\bibnamefont {Cui}}, \ and\
  \bibinfo {author} {\bibfnamefont {W.}~\bibnamefont {Chen}},\ } {\bibfield  {journal} {\bibinfo  {journal} {The Astrophysical
  Journal}\ }\textbf {\bibinfo {volume} {482}},\ \bibinfo {pages} {L155}
  (\bibinfo {year} {1997})}\BibitemShut {NoStop}%
\bibitem [{\citenamefont {{Kong}}\ \emph {et~al.}(2014)\citenamefont {{Kong}},
  \citenamefont {{Li}},\ and\ \citenamefont {{Bambi}}}]{Kong14}%
  \BibitemOpen
  \bibfield  {author} {\bibinfo {author} {\bibfnamefont {L.}~\bibnamefont
  {{Kong}}}, \bibinfo {author} {\bibfnamefont {Z.}~\bibnamefont {{Li}}}, \ and\
  \bibinfo {author} {\bibfnamefont {C.}~\bibnamefont {{Bambi}}},\ } {\bibfield  {journal} {\bibinfo
  {journal} {Astrophys. J.}\ }\textbf {\bibinfo {volume} {797}},\ \bibinfo
  {eid} {78} (\bibinfo {year} {2014})},\  {arXiv:1405.1508 [gr-qc]} \BibitemShut
  {NoStop}%
\bibitem [{\citenamefont {{Punsly}}\ and\ \citenamefont
  {{Coroniti}}(1990)}]{ref85}%
  \BibitemOpen
  \bibfield  {author} {\bibinfo {author} {\bibfnamefont {B.}~\bibnamefont
  {{Punsly}}}\ and\ \bibinfo {author} {\bibfnamefont {F.~V.}\ \bibnamefont
  {{Coroniti}}},\ } {\bibfield  {journal}
  {\bibinfo  {journal} {\apj}\ }\textbf {\bibinfo {volume} {354}},\ \bibinfo
  {pages} {583} (\bibinfo {year} {1990})}\BibitemShut {NoStop}%
\bibitem [{\citenamefont {Koide}(2003)}]{ref87}%
  \BibitemOpen
  \bibfield  {author} {\bibinfo {author} {\bibfnamefont {S.}~\bibnamefont
  {Koide}},\ } {\bibfield  {journal}
  {\bibinfo  {journal} {Phys. Rev. D}\ }\textbf {\bibinfo {volume} {67}},\
  \bibinfo {pages} {104010} (\bibinfo {year} {2003})}\BibitemShut {NoStop}%
\bibitem [{\citenamefont {Tchekhovskoy}\ \emph {et~al.}(2010)\citenamefont
  {Tchekhovskoy}, \citenamefont {Narayan},\ and\ \citenamefont
  {McKinney}}]{ref89}%
  \BibitemOpen
  \bibfield  {author} {\bibinfo {author} {\bibfnamefont {A.}~\bibnamefont
  {Tchekhovskoy}}, \bibinfo {author} {\bibfnamefont {R.}~\bibnamefont
  {Narayan}}, \ and\ \bibinfo {author} {\bibfnamefont {J.~C.}\ \bibnamefont
  {McKinney}},\ } {\bibfield
  {journal} {\bibinfo  {journal} {Astrophys. J.}\ }\textbf {\bibinfo {volume}
  {711}},\ \bibinfo {pages} {50} (\bibinfo {year} {2010})},\  {arXiv:0911.2228 [astro-ph.HE]} \BibitemShut
  {NoStop}%
\bibitem [{\citenamefont {Camilloni}\ \emph {et~al.}(2022)\citenamefont
  {Camilloni}, \citenamefont {Dias}, \citenamefont {Grignani}, \citenamefont
  {Harmark}, \citenamefont {Oliveri}, \citenamefont {Orselli}, \citenamefont
  {Placidi},\ and\ \citenamefont {Santos}}]{Camilloni:2022kmx}%
  \BibitemOpen
  \bibfield  {author} {\bibinfo {author} {\bibfnamefont {F.}~\bibnamefont
  {Camilloni}}, \bibinfo {author} {\bibfnamefont {O.~J.~C.}\ \bibnamefont
  {Dias}}, \bibinfo {author} {\bibfnamefont {G.}~\bibnamefont {Grignani}},
  \bibinfo {author} {\bibfnamefont {T.}~\bibnamefont {Harmark}}, \bibinfo
  {author} {\bibfnamefont {R.}~\bibnamefont {Oliveri}}, \bibinfo {author}
  {\bibfnamefont {M.}~\bibnamefont {Orselli}}, \bibinfo {author} {\bibfnamefont
  {A.}~\bibnamefont {Placidi}}, \ and\ \bibinfo {author} {\bibfnamefont
  {J.~E.}\ \bibnamefont {Santos}},\ } {\bibfield  {journal} {\bibinfo  {journal}
  {JCAP}\ }\textbf {\bibinfo {volume} {07}},\ \bibinfo {pages} {032} (\bibinfo
  {year} {2022})},\  {arXiv:2201.11068
  [gr-qc]} \BibitemShut {NoStop}%
\bibitem [{\citenamefont {Camilloni}\ \emph {et~al.}(2023)\citenamefont
  {Camilloni}, \citenamefont {Harmark}, \citenamefont {Orselli},\ and\
  \citenamefont {Rodriguez}}]{Camilloni:2023wyn}%
  \BibitemOpen
  \bibfield  {author} {\bibinfo {author} {\bibfnamefont {F.}~\bibnamefont
  {Camilloni}}, \bibinfo {author} {\bibfnamefont {T.}~\bibnamefont {Harmark}},
  \bibinfo {author} {\bibfnamefont {M.}~\bibnamefont {Orselli}}, \ and\
  \bibinfo {author} {\bibfnamefont {M.~J.}\ \bibnamefont {Rodriguez}},\
  }\href@noop {} {\  (\bibinfo {year} {2023})},\  {arXiv:2307.06878 [gr-qc]} \BibitemShut
  {NoStop}%
\bibitem [{\citenamefont {Pei}\ \emph {et~al.}(2016)\citenamefont {Pei},
  \citenamefont {Nampalliwar}, \citenamefont {Bambi},\ and\ \citenamefont
  {Middleton}}]{ref90}%
  \BibitemOpen
  \bibfield  {author} {\bibinfo {author} {\bibfnamefont {G.}~\bibnamefont
  {Pei}}, \bibinfo {author} {\bibfnamefont {S.}~\bibnamefont {Nampalliwar}},
  \bibinfo {author} {\bibfnamefont {C.}~\bibnamefont {Bambi}}, \ and\ \bibinfo
  {author} {\bibfnamefont {M.~J.}\ \bibnamefont {Middleton}},\ } {\bibfield  {journal} {\bibinfo  {journal}
  {Eur. Phys. J. C}\ }\textbf {\bibinfo {volume} {76}},\ \bibinfo {pages} {534}
  (\bibinfo {year} {2016})},\
  {arXiv:1606.04643 [gr-qc]} \BibitemShut {NoStop}%
\bibitem{Fender1} R. P. Fender, T. Belloni, and E. Gallo, Mon. Not. Roy. Astron. Soc. {\bf355},  1105 (2004), arXiv: astro-ph/0409360.
\bibitem{Fender2} R. P. Fender, \textit{Jets from X-ray binaries, in Compact stellar X-ray sources}, edited by W. Lewin and M. van der Klis (Cambridge University Press, UK, 2006), pp. 381-419. arXiv: astro-ph/0303339.
\bibitem [{\citenamefont {{Middleton}}\ \emph {et~al.}(2014)\citenamefont
  {{Middleton}}, \citenamefont {{Miller-Jones}},\ and\ \citenamefont
  {{Fender}}}]{ref94}%
  \BibitemOpen
  \bibfield  {author} {\bibinfo {author} {\bibfnamefont {M.~J.}\ \bibnamefont
  {{Middleton}}}, \bibinfo {author} {\bibfnamefont {J.~C.~A.}\ \bibnamefont
  {{Miller-Jones}}}, \ and\ \bibinfo {author} {\bibfnamefont {R.~P.}\
  \bibnamefont {{Fender}}},\ } {\bibfield
  {journal} {\bibinfo  {journal} {mnras}\ }\textbf {\bibinfo {volume} {439}},\
  \bibinfo {pages} {1740} (\bibinfo {year} {2014})},\  {arXiv:1401.1829 [astro-ph.HE]} \BibitemShut
  {NoStop}%
\bibitem [{\citenamefont {Tchekhovskoy}\ \emph {et~al.}(2011)\citenamefont
  {Tchekhovskoy}, \citenamefont {Narayan},\ and\ \citenamefont
  {McKinney}}]{Tchekhovskoyetal.2011}%
  \BibitemOpen
  \bibfield  {author} {\bibinfo {author} {\bibfnamefont {A.}~\bibnamefont
  {Tchekhovskoy}}, \bibinfo {author} {\bibfnamefont {R.}~\bibnamefont
  {Narayan}}, \ and\ \bibinfo {author} {\bibfnamefont {J.~C.}\ \bibnamefont
  {McKinney}},\ } {\bibfield
  {journal} {\bibinfo  {journal} {Monthly Notices of the Royal Astronomical
  Society: Letters}\ }\textbf {\bibinfo {volume} {418}},\ \bibinfo {pages}
  {L79} (\bibinfo {year} {2011})}\BibitemShut {NoStop}%
\bibitem [{\citenamefont {Russell}\ \emph {et~al.}(2013)\citenamefont
  {Russell}, \citenamefont {Gallo},\ and\ \citenamefont
  {Fender}}]{Russell2013ws}%
  \BibitemOpen
  \bibfield  {author} {\bibinfo {author} {\bibfnamefont {D.~M.}\ \bibnamefont
  {Russell}}, \bibinfo {author} {\bibfnamefont {E.}~\bibnamefont {Gallo}}, \
  and\ \bibinfo {author} {\bibfnamefont {R.~P.}\ \bibnamefont {Fender}},\
  } {\bibfield  {journal} {\bibinfo
  {journal} {Mon. Not. Roy. Astron. Soc.}\ }\textbf {\bibinfo {volume} {431}},\
  \bibinfo {pages} {405} (\bibinfo {year} {2013})},\  {arXiv:1301.6771 [astro-ph.HE]} \BibitemShut
  {NoStop}%
\end{thebibliography}
%

\end{document}